\documentclass[journal]{IEEEtran}
\usepackage{cite}
\usepackage{amsmath,amssymb,amsfonts}
\usepackage{algorithmic}
\usepackage{graphicx}
\usepackage{textcomp}
\usepackage{wrapfig}

\usepackage{physics}
\usepackage{bm}
\usepackage{booktabs}
\usepackage{multirow}
\usepackage{subcaption}
\usepackage{balance}

\newcommand{\ee}{\mathrm{e}}
\newcommand{\jj}{\mathrm{j}}
\newcommand{\vct}[1]{\bm{#1}}
\makeatletter
\let\MYcaption\@makecaption
\makeatother
\usepackage[font=footnotesize]{subcaption}
\makeatletter
\let\@makecaption\MYcaption
\makeatother
\usepackage{caption}
\def\BibTeX{{\rm B\kern-.05em{\sc i\kern-.025em b}\kern-.08em
    T\kern-.1667em\lower.7ex\hbox{E}\kern-.125emX}}
\begin{document}

\title{Autofocus Method for Human-Body Imaging under Respiratory Motion Using Synthetic Aperture Radar}

\author{Masaya~Kato and Takuya~Sakamoto
\thanks{M.~Kato and T.~Sakamoto are with the Department of Electrical Engineering, Graduate School of Engineering, Kyoto University, Kyoto 615-8510, Japan.}
}
\markboth{}
{Kato and Sakamoto: Autofocus Method for Human-Body Imaging under Respiratory Motion Using Synthetic Aperture Radar}

\maketitle

\begin{abstract}
This study presents an effective autofocusing approach for synthetic aperture radar imaging of the human body under conditions of respiratory motion. The proposed method suppresses respiratory-motion-induced phase errors by separating radar echoes in the spatial- and time-frequency domains and estimating phase errors individually for each separated echo. By compensating for the estimated phase errors, synthetic aperture radar images focused on all scattering points are generated, even when multiple body parts exhibit different motions due to respiration. The performance of the proposed method is evaluated through experiments with four participants in the supine position. Compared with a conventional method, the proposed approach improves image quality by a factor of 5.1 in terms of Muller--Buffington sharpness, and reduces the root-mean-square error with respect to a reference point cloud from 34 mm to 20 mm.
\end{abstract}

\begin{IEEEkeywords}
 Classification, heartbeat features, individual identification, millimeter-wave radar, non-contact sensing
\end{IEEEkeywords}

\IEEEpeerreviewmaketitle

\section{Introduction}
\label{sec:introduction}
\IEEEPARstart{R}{adar}-based physiological signal measurement enables simple and non-contact monitoring of vital signs in humans by detecting body displacements associated with respiration and heartbeat~\cite{Resp Iwata,Resp Koda,Heartbeat Itsuki}. In practice, respiratory-induced displacement strongly affects the chest and abdomen, making accurate radar-based heartbeat measurement challenging for these regions. In comparison, body parts such as the limbs and head exhibit much smaller respiratory motion and provide more suitable locations for heartbeat detection. Therefore, appropriate selection of measurement locations on the human body is a critical issue in radar-based physiological signal measurement.

Conventional radar measurements using a fixed radar position and a limited antenna aperture may not provide sufficient spatial resolution to resolve detailed structures of the human body, making it difficult to associate measured signals with specific body parts. Although camera-assisted approaches have been proposed to address this limitation~\cite{Depth camera Konishi 1,Depth camera Konishi 2,Depth camera Koshisaka,Camera 1,Camera 2}, such approaches raise privacy concerns.
In this context, synthetic aperture radar (SAR) imaging, which has been widely used in conjunction with synthetic aperture processing to achieve high spatial resolution through relative motion between the radar and the target (e.g., via mechanical scanning of the radar platform), provides a framework for estimating the structural information of the human body using radar alone, thereby preserving privacy.

However, SAR imaging suffers from image degradation caused by phase errors induced by non-negligible target motion during aperture synthesis, making autofocusing techniques essential for practical imaging~\cite{Autofocus Review}.
Existing autofocusing methods can be broadly classified into parametric approaches~\cite{Keystone,Radon,Doppler,Tracking,bessel 1,bessel 2,FrFT} and non-parametric approaches~\cite{PGA,Sharpness,Entropy}. However, to the best of the authors' knowledge, autofocusing techniques that explicitly address respiratory motion in stationary humans have not been reported. Respiratory motion induces non-negligible phase errors in SAR imaging, and suppressing such motion is essential for high-precision imaging of the human body.

In this study, we propose an autofocusing method to compensate for respiratory-motion-induced phase errors.
Our proposed method employs a mixture model to represent echoes affected by respiratory motion, thereby enabling robust separation of reflections from multiple body parts exhibiting different respiratory motions.
This echo separation process adopts a parametric approach, whereas the subsequent phase error estimation is performed using a non-parametric image-quality-based autofocus method. By integrating these parametric and non-parametric components, our proposed approach achieves reliable phase compensation under complex interference conditions. We demonstrate the effectiveness of the proposed method through experiments with four participants, and support the results with quantitative evaluation using ideal received signals and scattering power distributions.

\section{Synthetic Aperture Radar Processing for Targets in Motion}
\subsection{Synthetic Aperture Radar Imaging}
Let us assume that a complex-valued radar signal $s(r,\theta,t)$ is obtained, where $r$ denotes range, $\theta$ denotes azimuth angle, and $t$ denotes slow time. When a frequency-modulated continuous-wave (FMCW) radar with an antenna array is employed, $s(r,\theta,t)$ can be obtained by applying discrete Fourier transforms with respect to the fast-time index and the array element index. In such a case, $s(r,\theta,t)$ represents the complex baseband signal. Under the narrowband approximation within each range bin, phase compensation can be expressed using the carrier wavelength $\lambda$.

The backprojection method is commonly used for SAR imaging and can be considered equivalent to delay-and-sum processing, in which signals obtained at each radar platform position are coherently combined to form an image $I(\bm{x})$ according to:
\begin{align}
	I(\bm{x}) = \int_{t'-\Delta t/2}^{t'+\Delta t/2} 
	s\left(r(\bm{x},t),\theta(\bm{x},t),t\right)
    \ee^{-\mathrm{j}\psi(\bm{x},t)}\,\dd t,
	\label{eq:SARArray}
\end{align}
where $\Delta t$ denotes the integration time interval and $t'$ represents the central time of the integration interval.
Here, $r(\bm{x},t)$ is the distance between position $\bm{x}$ and the radar antenna at time $t$, and $\theta(\bm{x},t)$ is the azimuth angle of $\bm{x}$ relative to the array baseline at time $t$.
Moreover, $\psi(\bm{x},t)$ denotes the phase compensation term accounting for radar motion and is defined as $\psi(\bm{x},t)=4\pi r(\bm{x},t)/\lambda$, where $\lambda$ is the carrier wavelength.

\subsection{Phase Compensation for Targets in Motion}
The calculations described in the previous subsection assumed a stationary target, whereas in this subsection we consider SAR processing for targets in motion. Note that the phase term $\psi(\bm{x},t)$ compensates only for radar motion and does not account for target motion, thereby resulting in blurred images. Consequently, the SAR image becomes defocused, leading to degraded imaging quality.

If an additional phase shift $\phi(t)$ induced by the target motion is compensated for, a focused image can be obtained according to:
\begin{align}
  I_\mathrm{c}(\bm{x},\phi)=\int_{t'-\Delta t/2}^{t'+\Delta t/2}
  s\!\left(r(\bm{x},t),\theta(\bm{x},t),t\right)
  \ee^{-\jj\left(\psi(\bm{x},t)+\phi(t)\right)}\,\mathrm{d}t,
	\label{eq:ImageComp}
\end{align}
This process is referred to as autofocusing. The target motion is generally unknown \emph{a priori}, and therefore the phase error $\phi(t)$ must be estimated, as discussed in the next subsection.


\subsection{Image Sharpness and Phase Optimization}
\label{subsec:OptImage}
In many autofocusing methods, phase compensation is achieved by optimizing an image quality metric with respect to the unknown phase function $\phi$. One such metric is Muller and Buffington (MB) sharpness~\cite{MBSharpness and Sharpness,MBSharpness}, which is defined as
\begin{align}
  \gamma_\mathrm{MB}[I(\bm{x})]=\frac{\iiint |I(\bm{x})|^4\,\mathrm{d}\bm{x}}{\left|\iiint |I(\bm{x})|^2\,\mathrm{d}\bm{x}\right|^2}.
  \label{eq:MBSharpness}
\end{align}

Using $\gamma_\mathrm{MB}[I(\bm{x})]$, the phase function $\phi$ can be estimated by solving the following optimization problem:
\begin{align}
  \hat{\phi} = \arg\max_{\phi} \gamma_\mathrm{MB}[I_\mathrm{c}(\bm{x},\phi)].
	\label{eq:OptImageMetrics}
\end{align}

The phase-error estimation method introduced in this subsection assumes either a single target within the image or multiple targets sharing an identical phase error. Therefore, it may not function appropriately when multiple targets with distinct phase errors exist. Thus, in Section~\ref{sec:Prop}, we propose an autofocus method capable of handling multiple targets with different phase errors.

\section{Spectrogram Modeling for Targets with Respiratory Motion}
When a scanning radar observes targets undergoing respiratory motion, the Doppler frequency of the reflected signals contains components arising from both the radar motion and the respiratory motion of the targets.
Consequently, the received signal consists of superimposed reflections from multiple parts of the human body, each associated with a distinct Doppler frequency. In this section, we model the spectrogram of the received signal. First, we analytically derive the Doppler frequencies of the reflected waves when a scanning radar observes targets exhibiting respiratory motion. Next, we propose a mixture-model-based approach to represent the spectrogram of the received signal.

\subsection{Doppler Frequency Modeling of Respiratory Motion}
\label{subsec:RespDoppler}
In this subsection, we describe the Doppler frequency of radar echoes when a scanning radar observes targets undergoing respiratory motion.
When a target exhibits respiratory motion, the distance $R(t)$ between the target and the antenna can be expressed as the sum of the distance $\bar{R}(t)$ between the antenna and the nominal target position and the respiratory displacement $d(t)$.
Specifically, $R(t)=\bar{R}(t)+d(t)$.

If the nominal target position is denoted by $\bm{x}_\mathrm{t}$ and the antenna position is given by $\bm{x}_\mathrm{a}(t)=t\,\vct{v}+\bm{x}_0$, the distance can be written as $\bar{R}(t)=\left\| \bm{x}_\mathrm{a}(t)-\bm{x}_\mathrm{t} \right\|$, where $\vct{v}$ denotes the constant scanning velocity vector of the antenna and $\bm{x}_0$ is the initial position of the antenna.

Furthermore, the respiratory displacement component $d(t)$ is assumed to be periodic and is modeled using a truncated Fourier series as
\begin{align}
d(t)=\sum_{n=0}^{K}
\Re\left[c_n \ee^{\jj \omega_\mathrm{r} nt}\right],
\label{eq:RespComponent}
\end{align}
where $c_n \in \mathbb{C}$ denotes a complex coefficient associated with the $n$th respiratory harmonic, $\Re[\cdot]$ represents the real part of a complex number, and $\omega_\mathrm{r}=2\pi f_\mathrm{r}$ denotes the fundamental respiratory angular frequency.

The Doppler frequency $f_\mathrm{D}(t)$ is obtained by differentiating the
echo phase according to:
\begin{align}
  f_\mathrm{D}(t)
  &=\frac{2}{\lambda}
  \left\{\frac{\dd}{\dd t}\bar{R}(t)+\frac{\dd}{\dd t}d(t)\right\} \\
  &= \frac{2}{\lambda}
  \Biggl\{
  \frac{\vct{v}\cdot(t\vct{v}+\bm{x}_0-\bm{x}_\mathrm{t})}
       {\|t\vct{v}+\bm{x}_0-\bm{x}_\mathrm{t}\|}
  + \omega_\mathrm{r}\sum_{n=0}^{K}  n
    \Re\!\left[\jj c_n \ee^{\jj \omega_\mathrm{r} n t} \right]
  \Biggr\},
  \label{eq:DopplerResp2}
\end{align}
which indicates that the Doppler frequency consists of two components:
a radar-scanning-induced component and a respiration-induced component.

\subsection{Modeling of the Spectrogram of the Received Signal}
\label{subsec:Model}
Time--frequency analysis, such as the short-time Fourier transform (STFT), is commonly used to analyze time-varying Doppler frequencies, resulting in a spectrogram $S(t,f)$ that is a function of time $t$ and frequency $f$. In this study, the spectrogram is modeled as a superposition of multiple Gaussian-shaped components in the frequency domain:
\begin{align}
    S(t,f) = \sum_{m=1}^{M} \pi_m \, G(t,f \mid \omega_\mathrm{r}, \vct{c}_m),
\end{align}
where $M$ is the number of echoes, $\bm{\pi}=[\pi_1,\cdots,\pi_M]^\mathrm{T}$ denotes a vector of nonnegative weighting coefficients, and $\omega_\mathrm{r}$ and $C=[\vct{c}_1,\cdots,\vct{c}_M]$ are model parameters.

The function $G(t,f \mid \omega_\mathrm{r}, \vct{c}_m)$ is defined as
\begin{align}
  G(t,f \mid \omega_\mathrm{r}, \vct{c}_m)
  = \frac{1}{\Gamma}
    \exp\!\left(
      -\frac{(f-f_\mathrm{D}(t,\omega_\mathrm{r},\vct{c}_m))^2}{2\sigma^2}
    \right),
\end{align}
where $\Gamma$ is a normalization constant and $\sigma$ determines the frequency spread of each component in the spectrogram. The function $G$ attains its maximum at $f=f_\mathrm{D}(t,\omega_\mathrm{r},\vct{c}_m)$.

The Doppler trajectory of the $m$th component is modeled using the respiratory motion model introduced in the previous subsection, truncated to the first $K=2$ harmonics, as
\begin{align}
  f_\mathrm{D}(t,\omega_\mathrm{r},\vct{c}_m)
  = \sum_{n=0}^{2}
    \Re\!\left[c_n^{(m)} \ee^{\jj \omega_\mathrm{r} n t}\right]
    + c^{(m)} t,
  \label{eq:Model}
\end{align}
where a linear trend component $c^{(m)} t$ is included. The parameter vector is
\begin{align}
\hspace{-2cm}\vct{c}_m&=
\left[\Re(c_0^{(m)}),\Re(c_1^{(m)}),\Im(c_1^{(m)}),\right.\\
 &\hspace{2cm}\qquad \left.\Re(c_2^{(m)}),\Im(c_2^{(m)}),c^{(m)}\right]^\mathrm{T}
\end{align}
thus $\vct{c}_m \in \mathbb{R}^6$ collects the coefficients of the Doppler trajectory.

The function $f_\mathrm{D}(t,\omega_\mathrm{r},\bm{c}_m)$ is modeled as a sum of sinusoidal components (including up to the second harmonic) and a linear component, based on the Doppler frequency derived in Eq.~\eqref{eq:DopplerResp2}. Specifically, the sinusoidal terms represent Doppler frequency variations induced by respiratory motion, whereas the linear term accounts for the Doppler frequency component caused by radar scanning. In addition, the proposed model assumes that all radar echoes observed in the spectrogram originate from the same human body. Accordingly, the fundamental respiratory angular frequency is assumed to be identical and constant across all reflections, resulting in a common parameter $\omega_\mathrm{r}$ shared by all components. In this study, we empirically set $\sigma=0.3~\mathrm{Hz}$.

\section{Proposed Autofocusing Method}
\label{sec:Prop}
The image-quality-based autofocusing method described in Section~\ref{subsec:OptImage} enables estimation of a phase error for a single target in a SAR image. However, in human-body measurements, radar echoes from multiple body parts are superimposed, with each body part undergoing a distinct motion. As a result, multiple targets with different phase errors coexist within the image.

In this section, we propose an autofocusing method that individually estimates the phase errors of multiple targets and generates a SAR image focused on all targets. The proposed method consists of the following steps:
\begin{enumerate}
  \item Radar echoes from individual body parts are separated in the range--angle domain using the mixture model described in         Sections~\ref{subsec:RangeAngle} and~\ref{subsec:Model}.
  \item SAR images are generated for each separated target, and image-quality-based optimization (Section~\ref{subsec:OptImage}) is independently applied to estimate the phase error of each target.
  \item Using the estimated phase errors, focused images for individual body parts are obtained and subsequently integrated to form a SAR image focused on the entire human body.
\end{enumerate}

\subsection{Echo Separation in the Range-Angle Domain}
\label{subsec:RangeAngle}
This subsection describes a method for separating radar echoes originating from multiple body parts located at different ranges and angles by exploiting their spatial separability in the range--angle image. First, a radar image in the range--angle domain is constructed as
\begin{align}
	s_{\mathrm{p}}(r,\theta) &= \int_{t'-\Delta t/2}^{t'+\Delta t/2} \left| s\!\left(r,\theta,t\right) \right|^2 \dd t,
\end{align}
where the integration is performed over a temporal window centered at $t'$, and $s_{\mathrm{p}}(r,\theta)$ represents the accumulated signal power at each range--angle bin, which enhances dominant scattering components while suppressing temporal fluctuations.

From this image, local maxima whose amplitudes exceed a threshold $s_{\mathrm{th}}$ are extracted to form a set $P$ defined as
\begin{align}
P=\left\{(r,\theta)\middle|\, 
s_{\mathrm{p}}\geq s_{\mathrm{th}},\ 
\nabla s_{\mathrm{p}}=\bm{0},\ 
\nabla^2 s_{\mathrm{p}}< 0
\right\}.
\end{align}
Here, the condition $\nabla^2 s_{\mathrm{p}}<0$ indicates that the Hessian matrix $\nabla^2 s_{\mathrm{p}}$ is negative definite, ensuring that each element of $P$ corresponds to a local maximum.
Let $N$ denote the number of elements in $P$, and let the $n$-th element be represented as $(r_n,\theta_n)$ $(n=1,\ldots,N)$.

Assuming that each local maximum provides an estimate of a target position, we separate the reflected signals from individual targets by extracting spatial neighborhoods surrounding each maximum. The weighting function $w_n(r,\theta)$ for extracting the region around the $n$-th local maximum is defined as follows:
\begin{align}
w_n(r,\theta)=
\frac{
\exp\!\left(-\frac{(r-r_n)^2}{2\sigma_\mathrm{r}^2}\right)
\exp\!\left(-\frac{(\theta-\theta_n)^2}{2\sigma_\mathrm{a}^2}\right)
}{
\sum_{n'=1}^{N}
\exp\!\left(-\frac{(r-r_{n'})^2}{2\sigma_\mathrm{r}^2}\right)
\exp\!\left(-\frac{(\theta-\theta_{n'})^2}{2\sigma_\mathrm{a}^2}\right)
},
\end{align}
where $\sigma_\mathrm{r}$ and $\sigma_\mathrm{a}$ represent the standard deviations of the Gaussian distributions used for the weighting function. 
This normalization ensures that $\sum_{n=1}^{N} w_n(r,\theta)=1$ for any $(r,\theta)$, thereby enabling soft separation of overlapping reflections.
In this study, we set $\sigma_\mathrm{r} = 0.02~\mathrm{m}$ and $\sigma_\mathrm{a}=6.4^\circ$, based on the radar's range and angular resolutions.
The separated complex radar image $ s_{n}(r,\theta,t) $ is then obtained as
\begin{align}
  s_{n}\left(r,\theta,t\right)=w_n(r,\theta)s\left(r,\theta,t\right)
\end{align}

\begin{figure}[tb]
  \centering
  \begin{minipage}{0.49\columnwidth}
    \centering
    \includegraphics[width=\linewidth,pagebox=cropbox,clip]{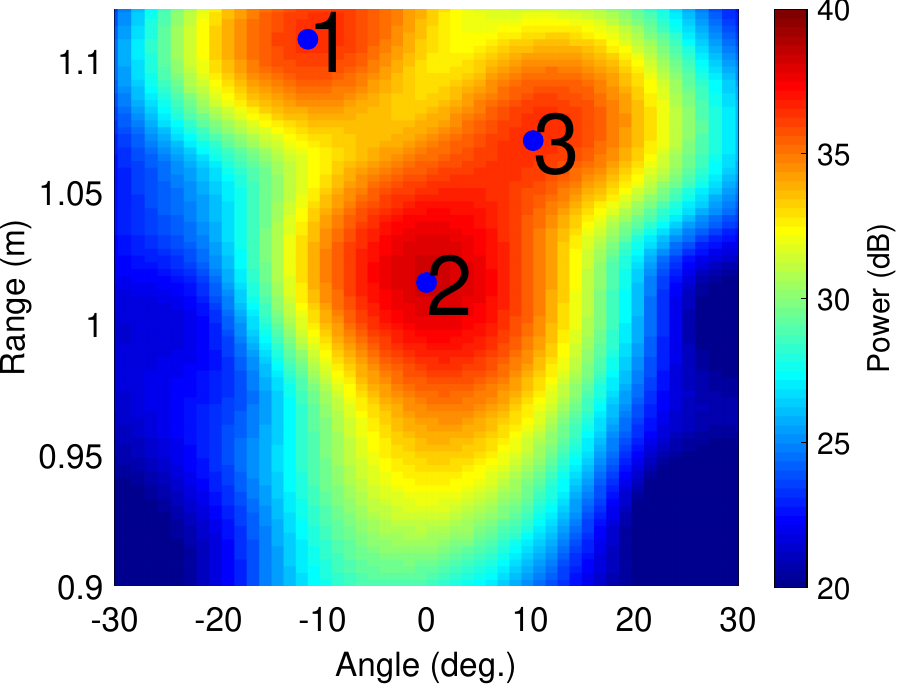}
    \subcaption{Measured radar image and three local maxima.}
  \end{minipage}
  \hfill
  \begin{minipage}{0.49\columnwidth}
    \centering
    \includegraphics[width=\linewidth,pagebox=cropbox,clip]{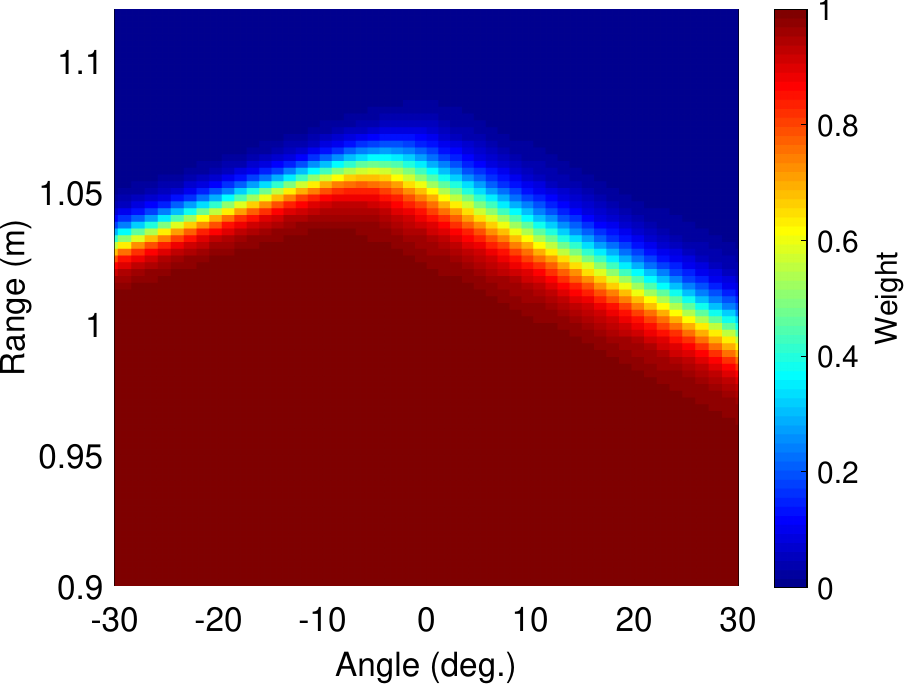}
    \subcaption{Weighting function in the range--angle domain.} 
  \end{minipage}
  \caption{Signal separation in the range--angle domain.}
  \label{fig:RangeAngle}
\end{figure}

Figs.~\ref{fig:RangeAngle}(a) and \ref{fig:RangeAngle}(b) illustrate examples of a radar image in the range--angle domain and the corresponding weighting function used to separate local maxima, respectively. These were obtained from the experiments described later.
The radar image clearly exhibits three local maxima, and the weighting function effectively extracts the region around the second local maximum. 
By using this process, reflected signals originating from multiple targets located at different ranges and angles are successfully separated.

\subsection{Echo Separation in the Time--Frequency Domain}
\label{subsec:Model2}
In the processing described in the previous subsection, echoes from multiple targets located at the same range and angle cannot be separated. 
To address this limitation, further separation is performed in the time--frequency domain by exploiting differences in the Doppler frequencies of targets sharing the same range and angle.

For each local maximum $(r_n,\theta_n)$, we extract the time series
$s_n(t)=s_n\left(r_n,\theta_n,t\right)$.
By applying the STFT to $s_n(t)$, we obtain its time--frequency representation $S_n(t,f)$.
Based on this representation, we estimate the parameters $\bm{\pi}^{(n)}$, $\omega_\mathrm{r}^{(n)}$, $\bm{c}^{(n)}$, and the mixture number $M^{(n)}$ of the mixture model introduced in Subsection~\ref{subsec:Model}.
Parameter estimation is performed by maximum likelihood estimation using the expectation--maximization algorithm.

To determine the number of mixture components, we employ the following information criterion:
\begin{align}
  I_\mathrm{MBIC} = -2L + \alpha N_\mathrm{p}\ln(N_\mathrm{s}),
  \label{eq:MBIC}
\end{align}
where $L$ denotes the log-likelihood, $N_\mathrm{s}$ is the number of samples, and $N_\mathrm{p}$ is the number of model parameters.
This criterion is a modified version of the Bayesian information criterion (BIC), in which the contribution of the penalty term is controlled by the coefficient $\alpha$. In this study, $\alpha=32$ was selected empirically. Preliminary evaluations revealed the standard BIC to be overly sensitive to correlated time--frequency samples produced by the STFT, which motivated our use of a modified penalty term in this study. Furthermore, assuming that at most two targets exist at the same range and angle, the maximum number of mixture components was set to $M^{(n)}=2$.

\begin{figure}[tb]
  \centering
  \begin{minipage}{0.49\columnwidth}
    \centering
    \includegraphics[width=\linewidth,pagebox=cropbox,clip]{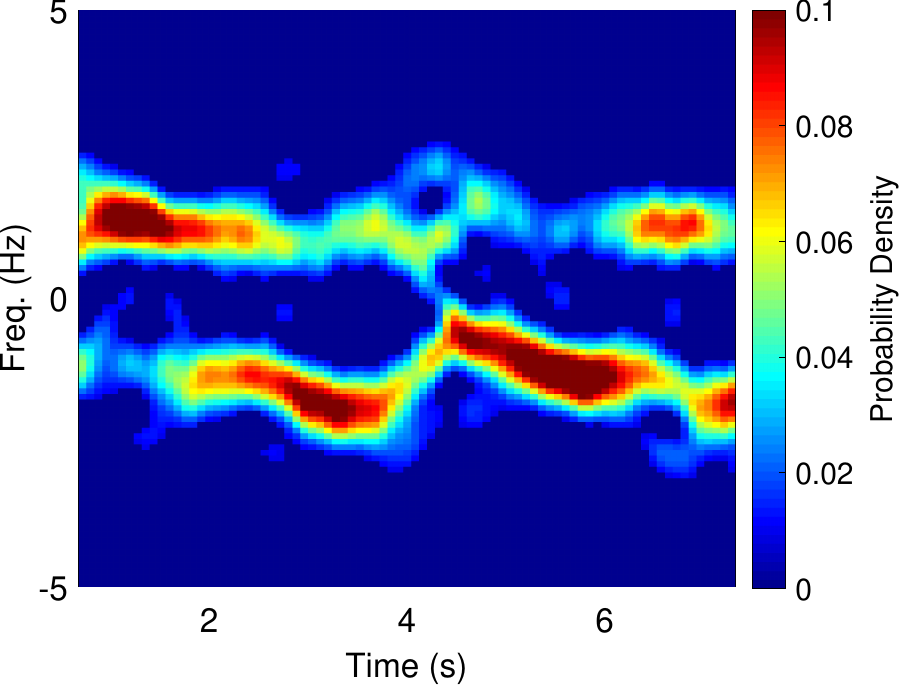}
    \subcaption{Measured spectrogram.}
  \end{minipage}
  \hfill
  \begin{minipage}{0.49\columnwidth}
    \centering
    \includegraphics[width=\linewidth,pagebox=cropbox,clip]{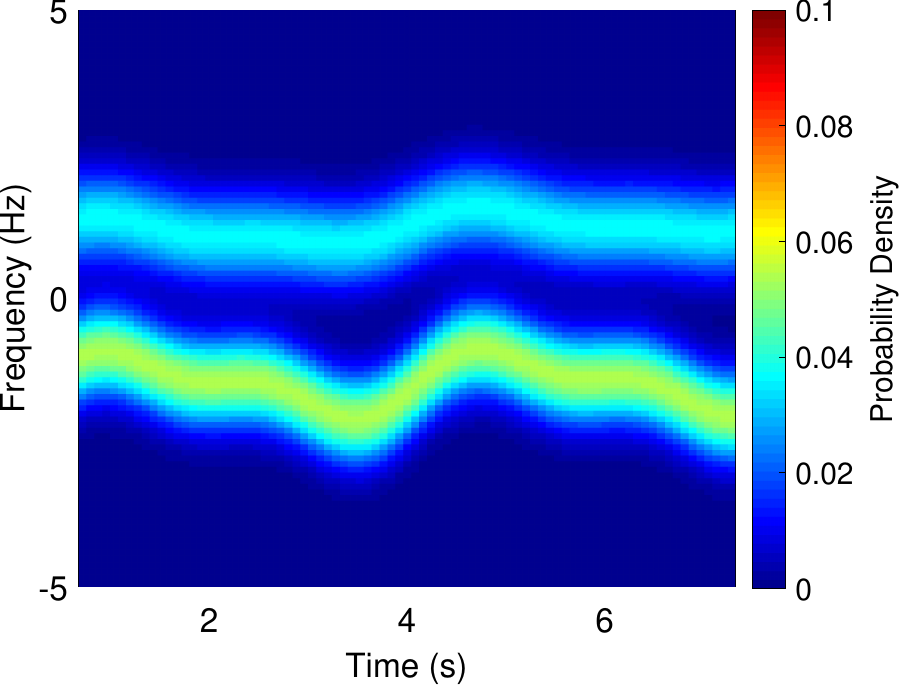}
\subcaption{Estimated spectrogram model.}
  \end{minipage}

  \medskip

  \begin{minipage}{0.49\columnwidth}
    \centering
    \includegraphics[width=\linewidth,pagebox=cropbox,clip]{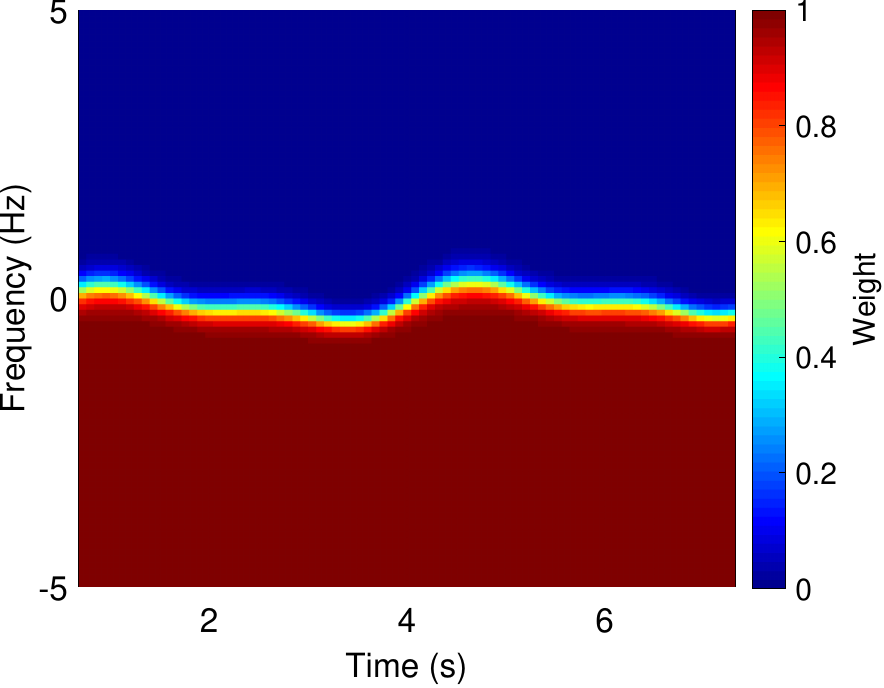}
\subcaption{Weighting function in the time--frequency domain.}
  \end{minipage}
  \caption{Signal separation in the time--frequency domain.}
  \label{fig:TFexamples}
\end{figure}

The echo corresponding to the $m$th component in the time--frequency domain for the $n$th component in the range--angle domain is referred to as the $m$th echo at the $n$th position.
We define a weighting function
\begin{align}
	w_{n,m}(t,f)=\frac{\pi_m^{(n)}G(t,f|\omega_\mathrm{r}^{(n)}, \bm{c}_m^{(n)})}{\displaystyle\sum_{m'=1}^{M^{(n)}}\pi_{m'}^{(n)}G(t,f|\omega_\mathrm{r}^{(n)}, \bm{c}_{m'}^{(n)})}
\end{align}
in the time--frequency domain to extract the $m$th echo at the $n$th position.
By using $w_{n,m}(t,f)$, we obtain
\begin{align}
	s_{n,m}\left(r,\theta,t\right)
	=
	\mathcal{F}_\mathrm{ST}^{-1}\!\left[
	w_{n,m}(t,f)\,
	\mathcal{F}_\mathrm{ST}\!\left[s_{n}\left(r,\theta,t\right)\right]
	\right],
\end{align}
where $\mathcal{F}_\mathrm{ST}$ and $\mathcal{F}_\mathrm{ST}^{-1}$ denote the STFT and its inverse, respectively.
With this approach, echoes from multiple body parts can be separated in the time--frequency domain, even when they are located at the same range and angle.

Figs.~\ref{fig:TFexamples}(a), \ref{fig:TFexamples}(b), and \ref{fig:TFexamples}(c) illustrate examples of the measured spectrogram, the estimated spectrogram based on the mixture model, and the corresponding weighting function, respectively.
These figures show that two components are detected in the time--frequency domain and that the weighting function selectively extracts one of the components.

\subsection{Image Generation Method}
\label{subsec:Image}
Synthetic aperture processing is applied to each separated signal $s_{n,m}(r,\theta,t)$ according to Eq.~(\ref{eq:ImageComp}), as described in Section~II.
For each $(n,m)$ pair, a SAR image is generated by estimating the residual phase using Eq.~(\ref{eq:OptImageMetrics}), where the phase $\hat{\phi}$ is determined so as to maximize the MB sharpness metric.
Because the phase compensation is performed individually for each body part, the resulting SAR images are well focused.

\begin{figure}[tb]
  \centering
  \begin{minipage}{0.80\columnwidth}
    \centering
    \includegraphics[width=\linewidth,pagebox=cropbox,clip]{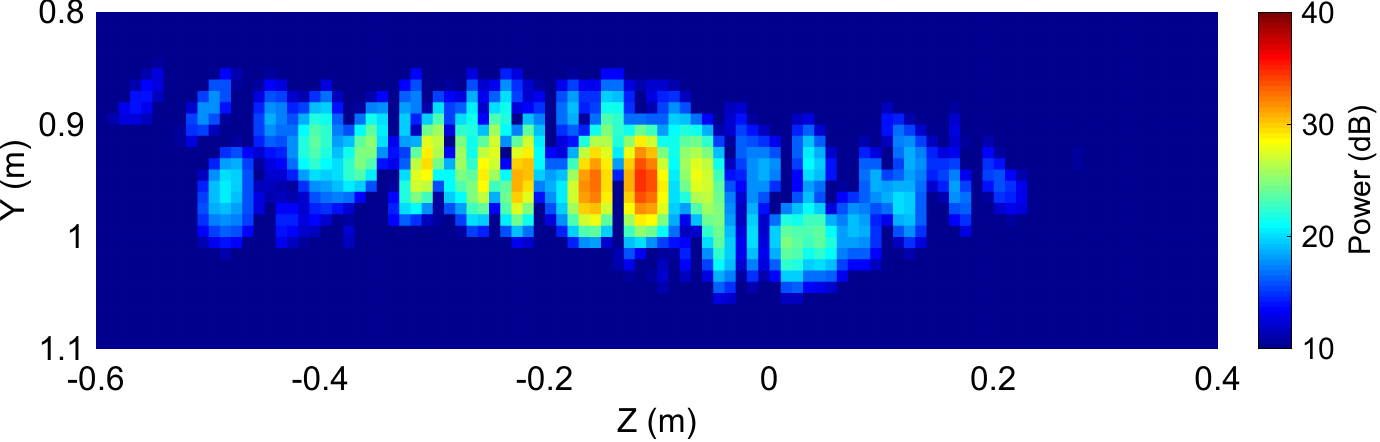}
    \subcaption{Image before phase compensation.}
  \end{minipage}
  \medskip
  \begin{minipage}{0.80\columnwidth}
    \centering
    \includegraphics[width=\linewidth,pagebox=cropbox,clip]{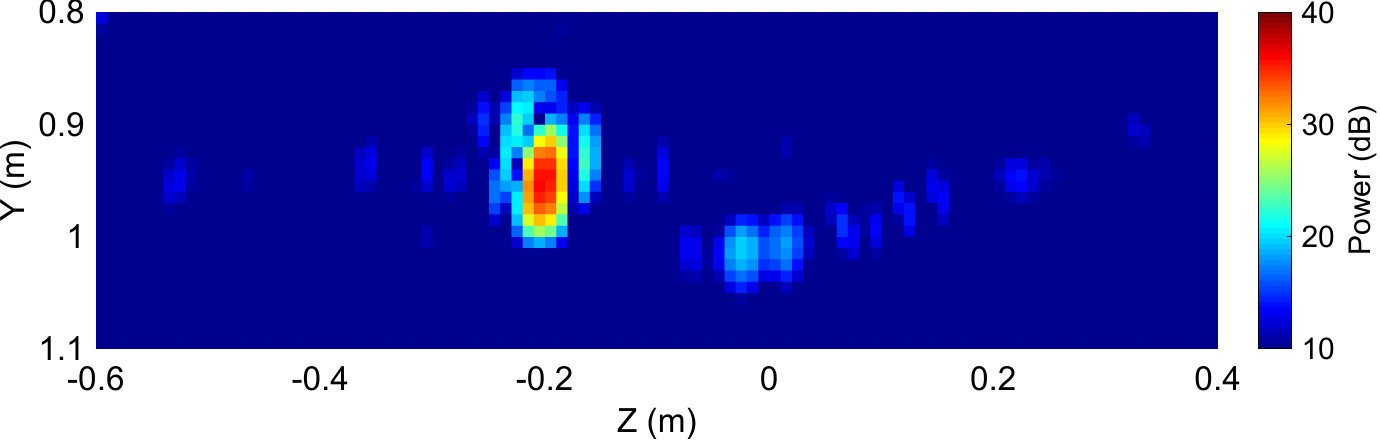}
    \subcaption{Image after phase compensation.}
  \end{minipage}
  \caption{Examples of SAR images before and after phase compensation ($n=70$, $p=1$, $m=1$).}
  \label{fig:hoseiEx}
\end{figure}

Finally, the focused SAR images for all $(n,m)$ pairs are combined to form a final image, with the images being summed incoherently as
\begin{align}
  I_\mathrm{p}(\bm{x})=\sum_{n=1}^{N}\sum_{m=1}^{M^{(n)}}\left|\hat{I}_{n,m}(\bm{x})\right|^2.
\end{align}
Figs.~\ref{fig:hoseiEx}(a) and \ref{fig:hoseiEx}(b) respectively illustrate examples of SAR images before and after phase compensation, as obtained from the experiments described later.
While the uncorrected image exhibits significant defocusing, the phase-compensated image shows well-focused target responses.

\section{Experimental Performance Evaluation}
\subsection{Details of the Radar Measurement Experiment}
We conducted radar measurement experiments with four participants to quantitatively evaluate the performance of our proposed autofocus method.
A millimeter-wave FMCW array radar operating at a center frequency of $79.0~\mathrm{GHz}$ with a bandwidth of $3.634~\mathrm{GHz}$ was used for these experiments.
The specifications of the radar system are summarized in Table~\ref{tab:ExpSyogen}.

The antenna configuration consists of a multiple-input multiple-output (MIMO) array with six transmitting and eight receiving elements, which can be virtually approximated as a two-dimensional array with 48 elements.
Both transmitting and receiving elements are spaced $1.9~\mathrm{mm}$ apart, corresponding to half the wavelength.
In this study, only the one-dimensional array formed by the first transmitting element and the eight receiving elements is analyzed.

In general, increasing the number of antenna elements leads to higher system complexity and cost.
Here, we demonstrate that highly accurate imaging can be achieved using only eight receiving elements, highlighting the effectiveness of the proposed method, even with a compact array configuration.

\begin{figure}[tb]
  \centering
  \includegraphics[width=0.7\columnwidth,pagebox=cropbox,clip]{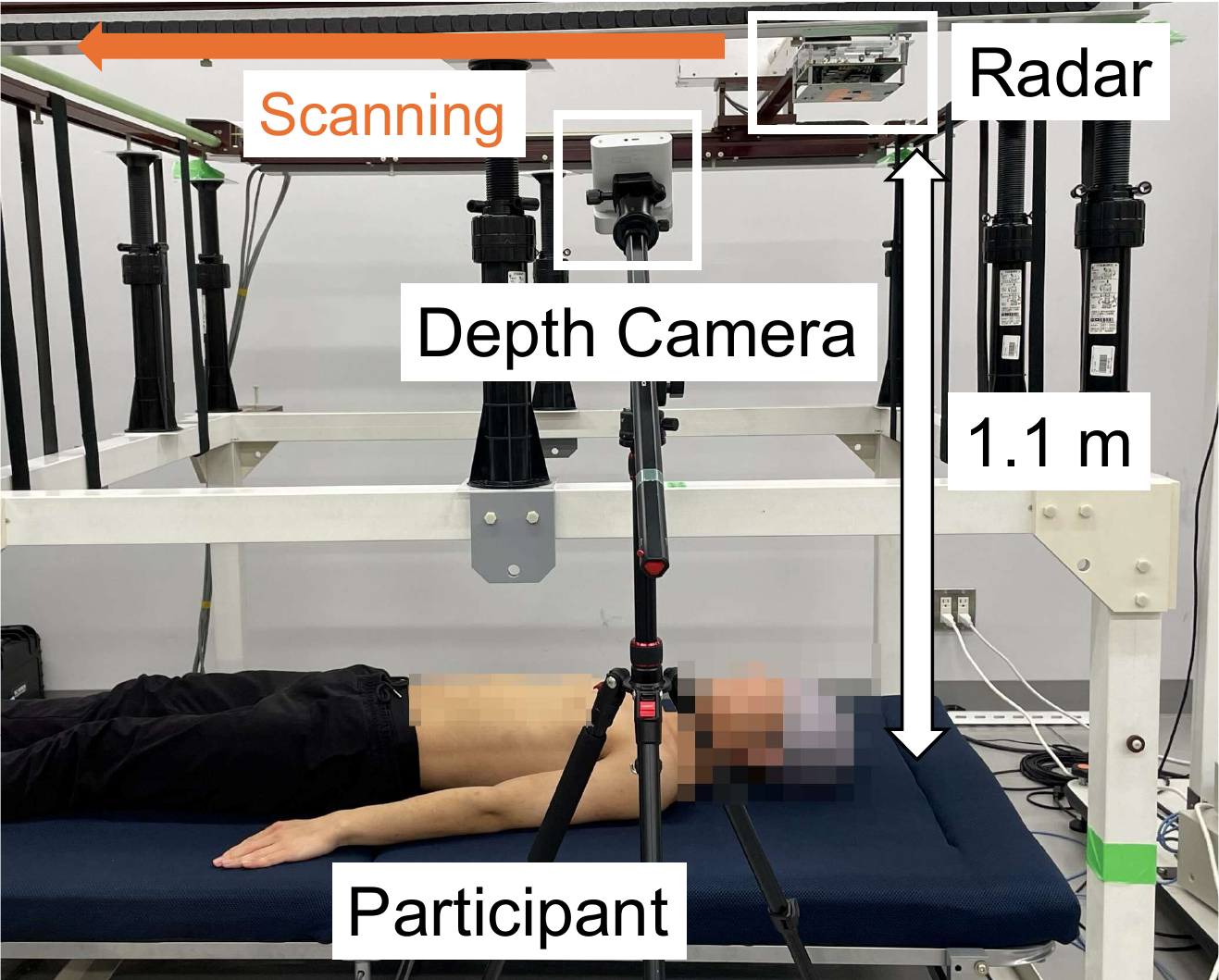}
  \caption{Photograph of the experimental setup for the radar measurements.}
  \label{fig:Experiment}
\end{figure}

The measurement setup for the radar experiment is shown in Fig.~\ref{fig:Experiment}.
The coordinate system was defined such that the array baseline direction corresponded to the $x$-axis, while the radar scanning direction corresponded to the negative $z$-axis.
In this coordinate system, the radar is mechanically scanned along the $z$-axis from $z=-0.45~\mathrm{m}$ to $z=0.39~\mathrm{m}$ at a constant scanning velocity of $|v|=9.9~\mathrm{mm/s}$, resulting in a total acquisition time of $85~\mathrm{s}$.
The mechanical scanning was performed using a mechanical scanner 
(DW3470AV1/O-13, DEVICE Co., Ltd., Kounosu, Saitama, Japan).

The upper surface of the bed was located at $y=1.1~\mathrm{m}$, and each participant lay on the bed with their body axis aligned along the $z$-axis. Two measurements were conducted for each participant.
To assess imaging accuracy, a depth camera (Azure Kinect DK, Microsoft Corp., Redmond, WA, USA) was additionally used to acquire a time-series depth map $z_\mathrm{ref}(x,y,t)$.

In this study, the received signals were segmented into temporal windows to generate SAR images.
For each measurement, SAR images were generated using a window length of $\Delta t = 8~\mathrm{s}$ over the total measurement duration of $85~\mathrm{s}$. By applying an overlap of $7.2~\mathrm{s}$ between consecutive windows, a total of $97$ images were obtained per measurement.
The imaging performance of our proposed autofocus method was evaluated by comparing its results with those obtained using the conventional method without autofocus.

\begin{table}[tb]
  \centering
  \caption{Specifications of the radar system used in the experiment}
  \label{tab:ExpSyogen}
  \begin{tabular}{lc}
      \toprule
      Parameter & Value \\ \midrule
      Modulation & FMCW \\
      Center frequency & $79.0~\mathrm{GHz}$ \\
      Bandwidth & $3.634~\mathrm{GHz}$ \\
      Number of transmit antennas & $6$ \\
      Number of receive antennas & $8$ \\
      Transmit element spacing & $1.9~\mathrm{mm}$ \\
      Receive element spacing & $1.9~\mathrm{mm}$ \\
      Sampling frequency & $25~\mathrm{Hz}$ \\
      \bottomrule
  \end{tabular}
\end{table}

\subsection{Reference Data Used for Performance Evaluation}
The following reference data were used to evaluate the imaging performance of the proposed autofocus method:
\begin{itemize}
  \item \textbf{Reference shape}\\
  A reference shape was obtained by converting the time-averaged depth data acquired by the depth camera into the point cloud $P_\mathrm{ref}$.
  \item \textbf{Reference radar image}\\
  A reference SAR image was generated using a physical optics approximation-based simulation~\cite{Sumi} based on the reference shape $P_\mathrm{ref}$.
  \item \textbf{Reference scattering centers}\\
  Reference scattering centers were estimated from the reference shape $P_\mathrm{ref}$ using the physical optics approximation-based method described in~\cite{Visualization}.
\end{itemize}
Respiratory-induced motion is effectively suppressed because the reference shape is derived from time-averaged depth camera data.
Therefore, these reference data are suitable for evaluating the imaging performance of our proposed method with respect to the static human body shape without the influence of respiratory motion.

\subsection{Performance Evaluation Metrics}
We quantitatively evaluated the imaging performance using metrics corresponding to the three types of reference data described in the previous subsection.
Specifically, image focusing accuracy was evaluated using the reference radar image, while the accuracy of estimated scattering points was evaluated using the reference shape and the reference scattering centers.
Imaging performance was quantitatively evaluated in terms of image focusing accuracy and the accuracy of the estimated scattering points.
The image focusing performance was evaluated using the MB sharpness metric and the correlation coefficient between the generated and reference radar images.

For each SAR image, local maxima exceeding a power threshold $I_\mathrm{th}$ were extracted and integrated to form a set of estimated scattering points,
\[
P_\mathrm{est}=\{\bm{p}_i^{(\mathrm{est})}\in\mathbb{R}^3 \mid i=1,\ldots,N_\mathrm{est}\}.
\]
The accuracy of the estimated scattering points was evaluated using the root mean squared (RMS) error defined as
\begin{align}
E_\mathrm{RMSE}
=
\sqrt{
\frac{1}{N_\mathrm{est}}
\sum_{i=1}^{N_\mathrm{est}}
\min_{\bm{p}\in P_\mathrm{ref}}
\left\|
\bm{p}_i^{(\mathrm{est})}-\bm{p}
\right\|^2
}.
\end{align}

\subsection{Performance Evaluation of the Proposed Method}
\label{sebsec:ExpEval}
\begin{figure}[tb]
  \centering
  \begin{minipage}{0.75\columnwidth} 
    \centering
    \includegraphics[width=\columnwidth,pagebox=cropbox,clip]{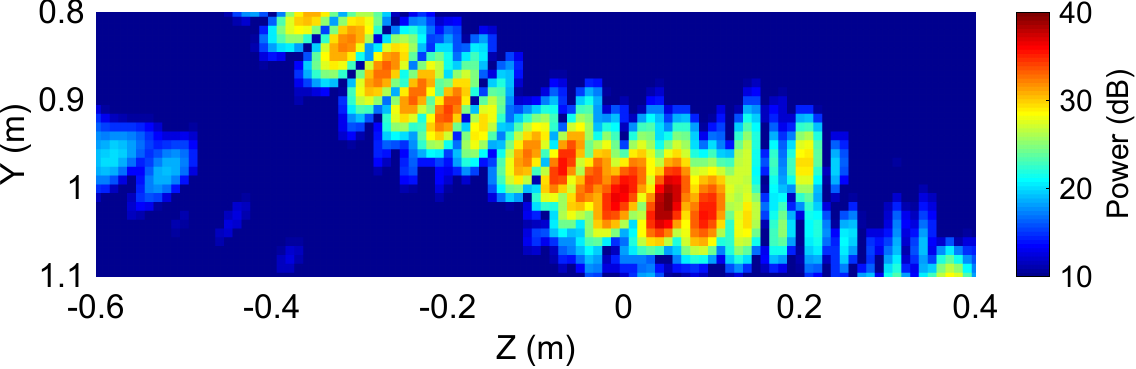}
    \subcaption{Conventional}
  \end{minipage}
  \begin{minipage}{0.75\columnwidth}
    \centering
    \includegraphics[width=\columnwidth,pagebox=cropbox,clip]{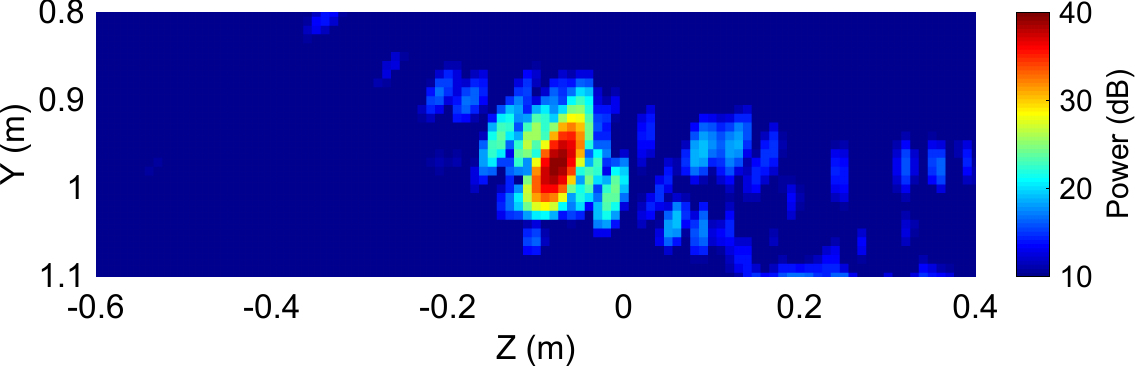}
    \subcaption{Proposed}
  \end{minipage}
  \begin{minipage}{0.75\columnwidth}
    \centering
    \includegraphics[width=\columnwidth,pagebox=cropbox,clip]{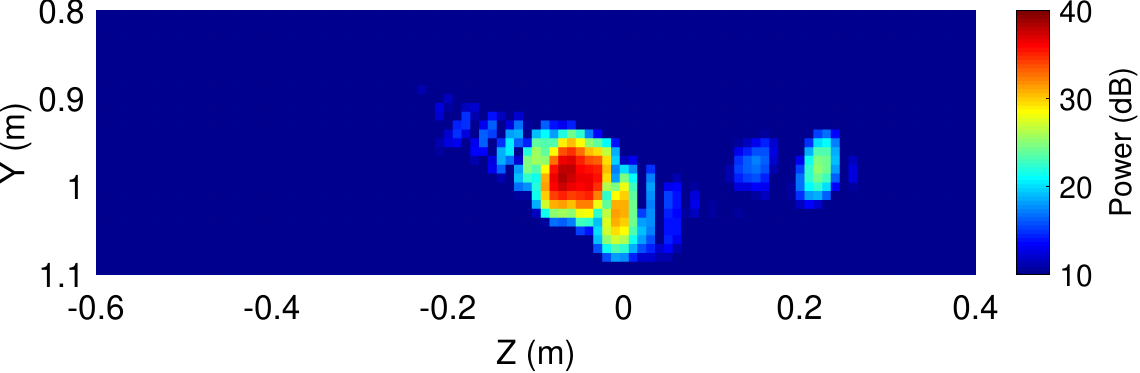}
    \subcaption{Reference}
  \end{minipage}
  \caption{Example of SAR image (participant 1, first measurement, chest region)}
  \label{fig:Imaging1-1}
\end{figure}

\begin{figure}[tb]
  \centering
  \begin{minipage}{0.75\columnwidth} 
    \centering
    \includegraphics[width=\columnwidth,pagebox=cropbox,clip]{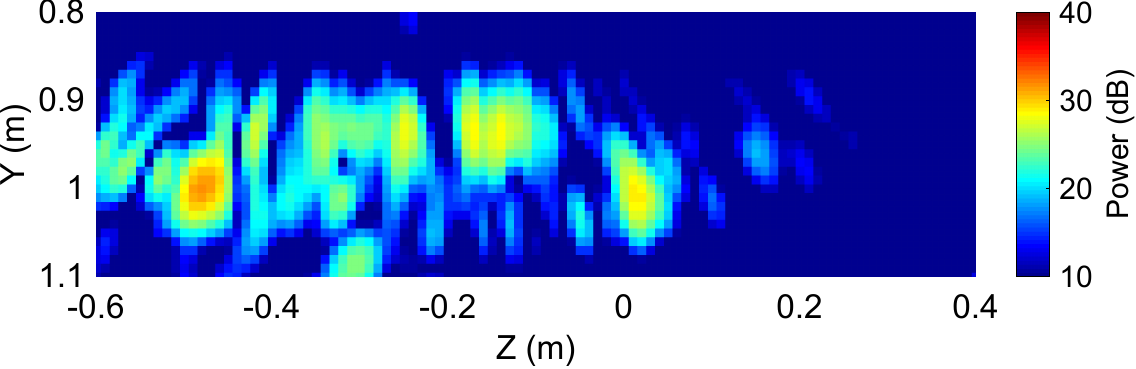}
    \subcaption{Conventional}
  \end{minipage}
  \begin{minipage}{0.75\columnwidth}
    \centering
    \includegraphics[width=\columnwidth,pagebox=cropbox,clip]{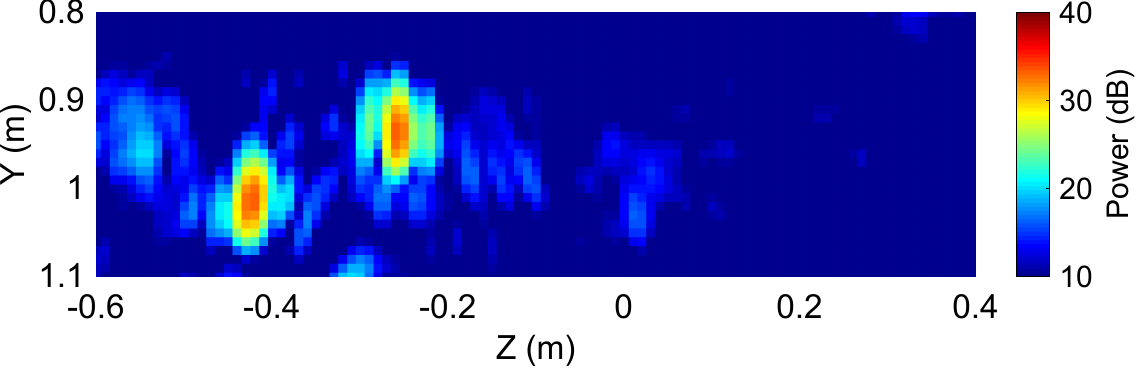}
    \subcaption{Proposed}
  \end{minipage}
  \begin{minipage}{0.75\columnwidth}
    \centering
    \includegraphics[width=\columnwidth,pagebox=cropbox,clip]{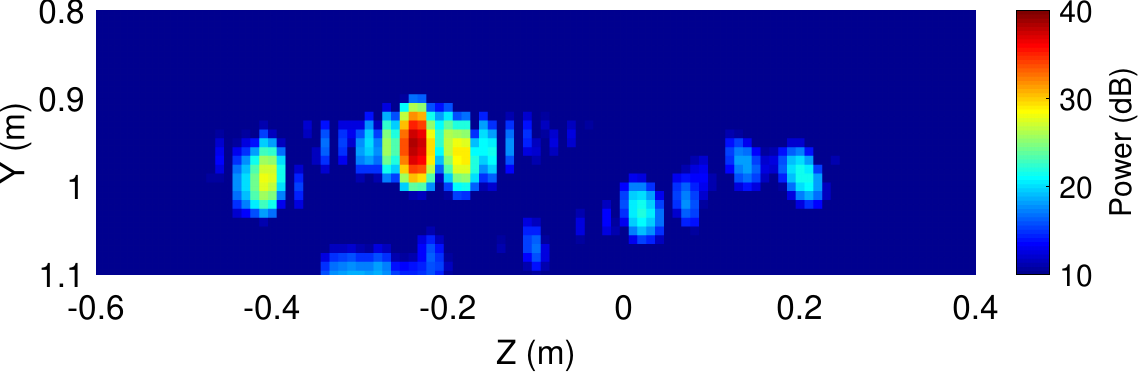}
    \subcaption{Reference}
  \end{minipage}
  \caption{Example of SAR image (participant 1, first measurement, abdomen region)}
  \label{fig:Imaging1-2}
\end{figure}

Examples of SAR images obtained from the experiments together with corresponding reference radar images are shown in Figs.~\ref{fig:Imaging1-1} and \ref{fig:Imaging1-2}.
The conventional method failed to produce well-focused images, making it difficult to identify individual scattering points.
In contrast, the proposed autofocus method yielded clearly focused images in which multiple distinct scattering points are observable, enabling accurate identification of their locations.
Comparisons with the reference radar images further demonstrate that the images produced using our proposed method are substantially closer to the ideal references than are those obtained by the conventional method.

The average MB sharpness values of the SAR images are summarized for all participants in Table~\ref{tab:ExpMB}.
For all measurements, the proposed method consistently achieved significantly higher MB sharpness values than did the conventional method.
On average, MB sharpness was improved by a factor of 5.1 compared with the conventional method, confirming the effectiveness of the proposed autofocus approach for enhancing image focus.

\begin{table}[tb]
  \centering
   \caption{Comparison of average MB sharpness values of SAR images between conventional and proposed methods}
  \begin{tabular}{cccc}
      \toprule
      \multirow{2}{*}{Participant} & \multirow{2}{*}{Measurement} & \multicolumn{2}{c}{MB sharpness (m$^{-3}$)} \\ \cmidrule(lr){3-4}
      &                       & Conventional & Proposed \\ \midrule
      \multirow{2}{*}{1} & 1st & 7.7$\times 10^{1}$ & 2.6$\times 10^{2}$ \\
                         & 2nd & 6.7$\times 10^{1}$ & 2.9$\times 10^{2}$ \\
      \multirow{2}{*}{2} & 1st & 6.8$\times 10^{1}$ & 3.2$\times 10^{2}$ \\
                         & 2nd & 7.7$\times 10^{1}$ & 3.3$\times 10^{2}$ \\
      \multirow{2}{*}{3} & 1st & 6.4$\times 10^{1}$ & 3.8$\times 10^{2}$ \\
                         & 2nd & 6.8$\times 10^{1}$ & 3.3$\times 10^{2}$ \\
      \multirow{2}{*}{4} & 1st & 4.6$\times 10^{1}$ & 3.4$\times 10^{2}$ \\
                         & 2nd & 5.5$\times 10^{1}$ & 3.0$\times 10^{2}$ \\ \cmidrule(lr){1-4}
      Average & & 6.5$\times 10^{1}$ & 3.2$\times 10^{2}$ \\ \bottomrule
  \end{tabular}
  \label{tab:ExpMB}
\end{table}

\begin{figure}[tb]
  \centering
  \begin{minipage}{0.75\columnwidth} 
    \centering
    \includegraphics[width=\columnwidth,pagebox=cropbox,clip]{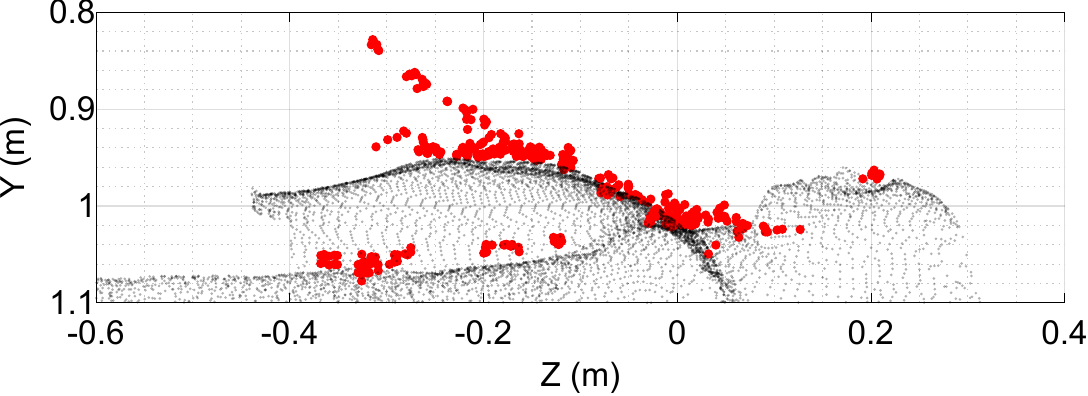}
    \subcaption{Conventional}
  \end{minipage}
  \begin{minipage}{0.75\columnwidth}
    \centering
    \includegraphics[width=\columnwidth,pagebox=cropbox,clip]{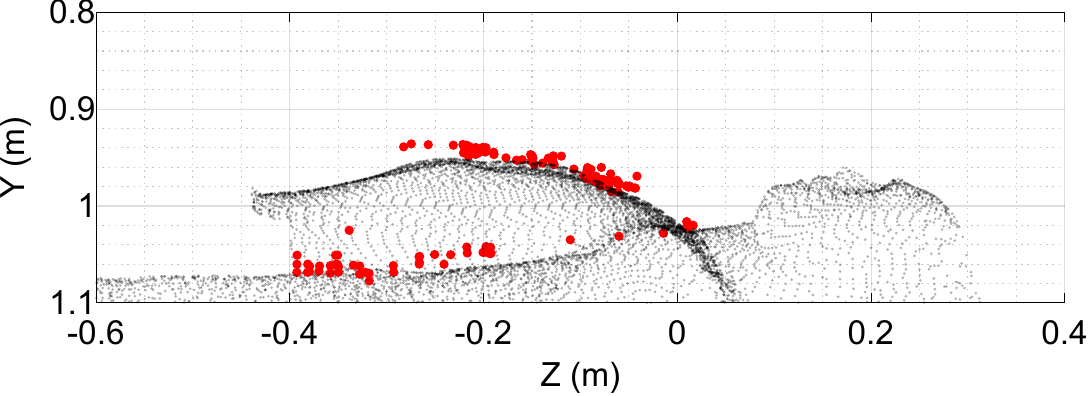}
    \subcaption{Proposed}
  \end{minipage}
  \begin{minipage}{0.75\columnwidth}
    \centering
    \includegraphics[width=\columnwidth,pagebox=cropbox,clip]{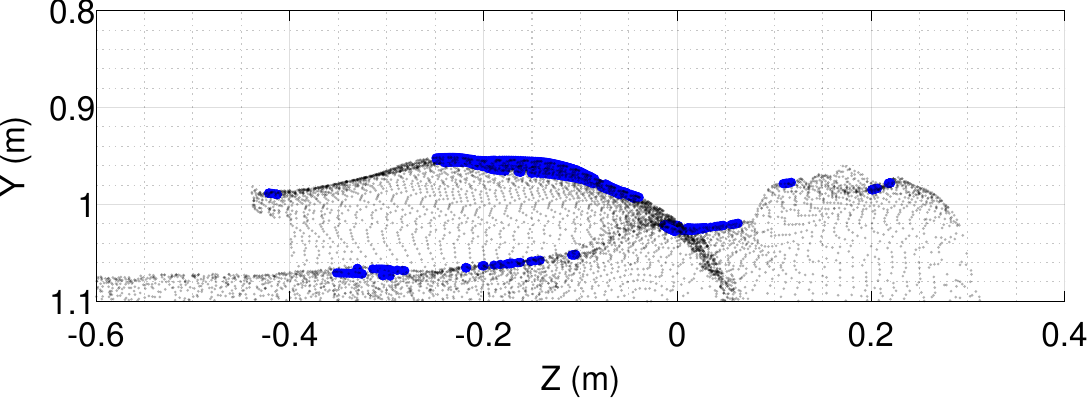}
    \subcaption{Reference}
  \end{minipage}
  \caption{Estimated scattering points (red), reference point cloud (black), and reference scattering points (blue) (participant 1, first measurement)}
  \label{fig:EstPoint1}
\end{figure}

Examples of estimated scattering points, reference scattering points, and reference point clouds are shown in Fig.~\ref{fig:EstPoint1}.
The estimated scattering points obtained using the conventional method are distributed away from the reference point cloud, indicating degraded localization accuracy.
In contrast, the proposed method produced scattering points that closely align with the reference point cloud.
Comparison with the reference scattering points reveals that the spatial distribution of the scattering points obtained using our proposed method closely matches that of the reference, whereas noticeable discrepancies can be observed for the conventional method.

However, it should be noted that scattering points around the head region were not successfully estimated by the proposed method.
This limitation arises because the phase optimization based on maximizing the MB sharpness emphasizes strong echoes while suppressing weaker ones, such as those originating from the head.

The RMS errors of the estimated scattering points for all participants are summarized in Table~\ref{tab:ExpRMSE}.
For every participant, the proposed method significantly reduced the RMS error compared with the conventional method.
On average, the RMS error was reduced from $34~\mathrm{mm}$ to $20~\mathrm{mm}$, corresponding to an improvement of $14~\mathrm{mm}$.

\begin{table}[tb]
  \centering
  \caption{RMS error between estimated scattering points and reference point cloud}
  \label{tab:ExpRMSE}
  \begin{tabular}{cccc}
      \toprule
      \multirow{2}{*}{Participant} & \multirow{2}{*}{Measurement} & \multicolumn{2}{c}{RMSE (mm)} \\ \cmidrule(lr){3-4}
                            &                       & Conventional & Proposed \\ \midrule
      \multirow{2}{*}{1} & 1st & 29 & 17 \\
                         & 2nd & 28 & 13 \\
      \multirow{2}{*}{2} & 1st & 29 & 18 \\
                         & 2nd & 26 & 21 \\
      \multirow{2}{*}{3} & 1st & 37 & 30 \\
                         & 2nd & 36 & 20 \\
      \multirow{2}{*}{4} & 1st & 44 & 17 \\
                         & 2nd & 45 & 24 \\ \cmidrule(lr){1-4}
      Average & & 34 & 20 \\ \bottomrule
  \end{tabular}
\end{table}

\section{Conclusion}
In this study, we propose an autofocusing method for SAR imaging of the human body under respiratory motion. Our proposed method separates echoes from multiple body parts that exhibit different respiratory motions and individually estimates their phase errors by combining range--angle separation, time--frequency-domain mixture modeling, and image-quality-based optimization. Integration of the individually focused images allows us to obtain a well-focused SAR image covering all targets.

Experimental validation was conducted with four participants imaged in a supine position.
Compared with the conventional method, our proposed approach achieved a 5.1-fold improvement in MB sharpness and reduced the RMS error of estimated scattering points from $34~\mathrm{mm}$ to $20~\mathrm{mm}$.
These results demonstrate that the proposed method effectively suppresses respiratory-motion-induced phase errors and enables high-precision human body imaging using SAR.

\section*{Ethics declarations}
This study involved human participants.
All ethical and experimental procedures were approved by the Ethics Committee of the Graduate School of Engineering, Kyoto University (Approval No.~202223), and informed consent was obtained from all participants.

\begin{IEEEbiography}[{\includegraphics[width=1in,height=1.25in,clip,keepaspectratio]{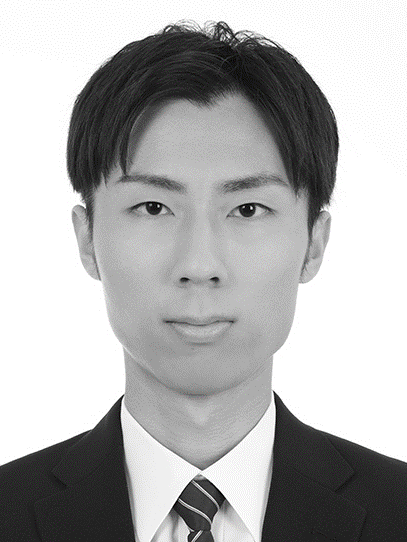}}]{Masaya Kato} received the B.E. degree in electrical and electronic engineering from Kyoto University, Kyoto, Japan, in 2023, and the M.E. degree in electrical engineering from the Graduate School of Engineering, Kyoto University, in 2025. His research interests include wireless human sensing, radar signal processing, and radar imaging.
\end{IEEEbiography}

\begin{IEEEbiography}[{\includegraphics[width=1in,height=1.25in,clip,keepaspectratio]{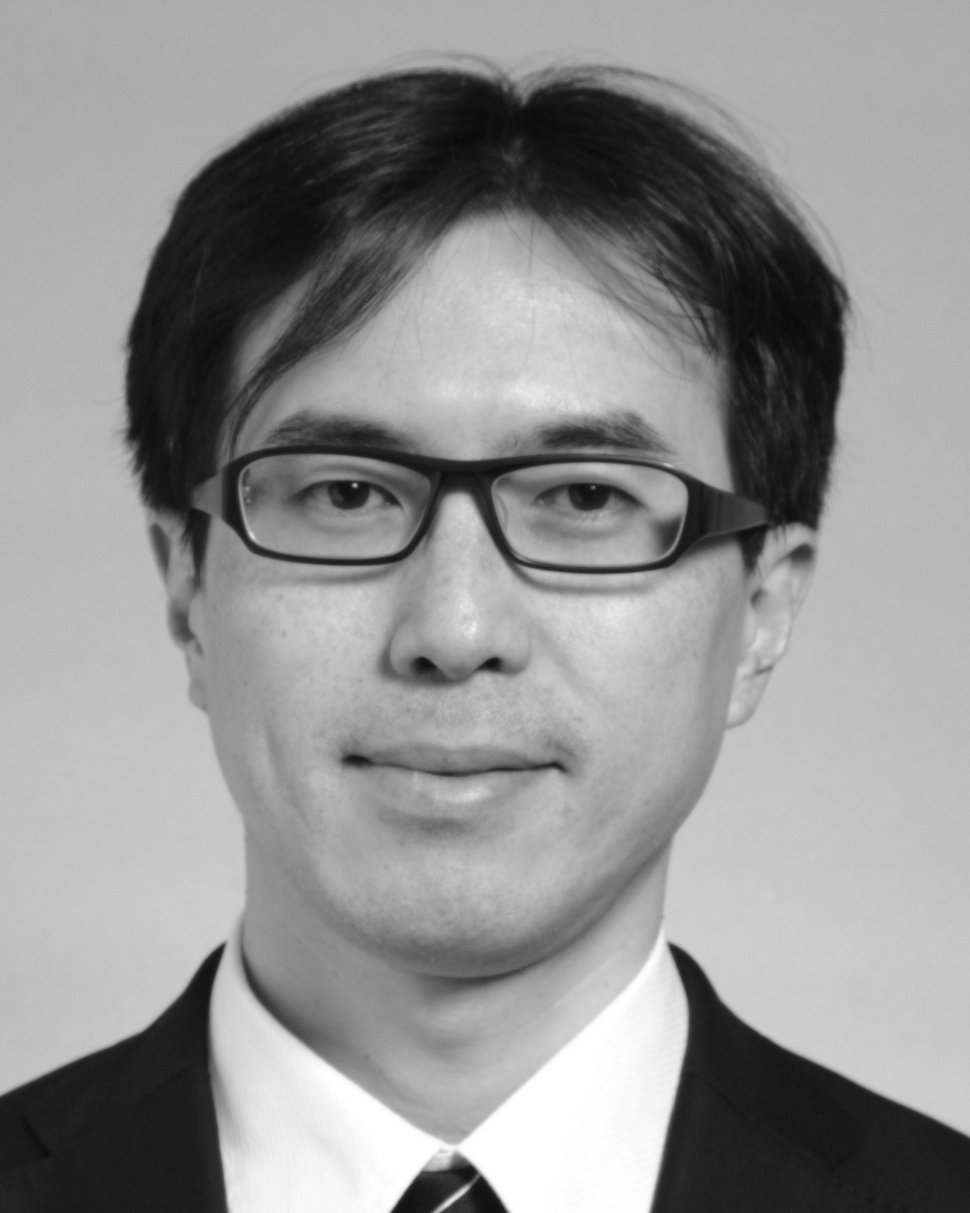}}]{Takuya Sakamoto} (Senior Member, IEEE) received the B.E. degree in Electrical and Electronic Engineering from Kyoto University, Kyoto, Japan, in 2000, and the M.I. and Ph.D. degrees in Communications and Computer Engineering from the Graduate School of Informatics, Kyoto University, in 2002 and 2005, respectively. He was a Visiting Researcher at Delft University of Technology, Delft, the Netherlands, from 2011 to 2013, and a Visiting Scholar at the University of Hawaii at Manoa, Honolulu, HI, USA, in 2017. From 2018 to 2022, he was also a PRESTO Researcher with the Japan Science and Technology Agency (JST), Japan. Since 2022, he has been a Professor with the Graduate School of Engineering, Kyoto University. His current research interests include wireless human sensing, radar signal processing, and radar-based measurement of physiological signals.

Prof. Sakamoto has received multiple Best Paper Awards from the International Symposium on Antennas and Propagation (ISAP), including awards in 2004 and 2012, the Masao Horiba Award in 2016, the Telecom System Technology Award from the Telecommunications Advancement Foundation in 2022, the Best Paper Award from the Institute of Electronics, Information and Communication Engineers (IEICE) Communications Society in 2023, and the Electronics Society Award from IEICE in 2025.
\end{IEEEbiography}


\begin{thebibliography}{22}

    \bibitem{Resp Iwata} S. Iwata, T. Koda, and T. Sakamoto, ``Multiradar data fusion for respiratory measurement of multiple people,'' \emph{IEEE Sensors Journal}, vol. 21, no. 22, pp. 25870--25879, Nov. 2021, DOI: 10.1109/JSEN.2021.3117707.

    \bibitem{Resp Koda} T. Koda, T. Sakamoto, S. Okumura, and H. Taki, ``Noncontact respiratory measurement for multiple people at arbitrary locations using array radar and respiratory-space clustering,'' \emph{IEEE Access}, vol. 9, pp. 106895--106906, Jul. 2021, DOI: 10.1109/ACCESS.2021.3099821.

    \bibitem{Heartbeat Itsuki}I. Iwata, T. Sakamoto, T. Matsumoto, and S. Hirata, ``Noncontact measurement of Heartbeat of Humans and Chimpanzees Using Millimeter-Wave Radar With Topology Method,'' \emph{IEEE Sensors Letters}, vol. 7, no.11, art. no. 7006104, Nov. 2023, DOI: 10.1109/ACCESS.2021.3099821.

    \bibitem{Depth camera Konishi 1} K. Konishi and T. Sakamoto, ``Automatic tracking of human body using millimeter-wave adaptive array radar for noncontact heart rate measurement,'' in \emph{Proc. 2018 Asia-Pacific Microwave Conference (APMC)}, Kyoto, Japan, Nov. 2018, pp.~836--838, DOI: 10.23919/APMC.2018.8617221.

    \bibitem{Depth camera Konishi 2} T. Sakamoto, K. Konishi, K. Yamashita, M. Muragaki, S. Okumura, and T. Sato, ``Adaptive array radar imaging of a human body for vital sign measurement,'' in \emph{Proc. 2018 IEEE International Symposium on Antennas and Propagation \& USNC/URSI National Radio Science Meeting}, Boston, MA, USA, Jul. 2018, pp.~617--618, DOI: 10.1109/APUSNCURSINRSM.2018.8608643.

    \bibitem{Depth camera Koshisaka} T. Koshisaka and T. Sakamoto, ``Radar-based pulse wave sensing at multiple sites using a 3D human model,'' \emph{IEEE Sensors Journal}, early access, Dec. 2024, DOI: 10.1109/JSEN.2024.3519571.

    \bibitem{Camera 1} I. V. Mikhelson, P. Lee, S. Bakhtiari, T. W. Elmer, A. K. Katsaggelos, and A. V. Sahakian, ``Noncontact millimeter-wave real-time detection and tracking of heart rate on an ambulatory subject,'' \emph{IEEE Transactions on Information Technology in Biomedicine}, vol. 16, no. 5, pp. 927--934, Sep. 2012, DOI: 10.1109/TITB.2012.2204760.

    \bibitem{Camera 2} A. Shokouhmand, S. Eckstrom, B. Gholami, and N. Tavassolian, ``Camera-augmented non-contact vital sign monitoring in real time,'' \emph{IEEE Sensors Journal}, vol. 22, no. 12, pp. 11965--11978, Jun. 2022, DOI: 10.1109/JSEN.2022.3172559.

    \bibitem{Autofocus Review} J. Chen, M. Xing, H. Yu, B. Liang, J. Peng, and G.-C. Sun, ``Motion compensation/autofocus in airborne synthetic aperture radar: a review,'' \emph{IEEE Geoscience and Remote Sensing Magazine}, vol. 10, no. 1, pp. 185--206, Mar. 2022, DOI: 10.1109/MGRS.2021.3113982.

    \bibitem{Keystone} C. Zeng, D. Li, X. Luo, D. Song, H. Liu, and J. Su, ``Ground maneuvering targets imaging for synthetic aperture radar based on second-order keystone transform and high-order motion parameter estimation,'' \emph{IEEE Journal of Selected Topics in Applied Earth Observations and Remote Sensing}, vol. 12, no. 11, pp. 4486--4501, Nov. 2019, DOI: 10.1109/JSTARS.2019.2951199.

    \bibitem{Radon} L. Lin, G. Sun, Z. Cheng, and Z. He, ``Long-time coherent integration for maneuvering target detection based on ITRT-MRFT,'' \emph{IEEE Sensors Journal}, vol. 20, no. 7, pp. 3718--3731, Apr. 2020, DOI: 10.1109/JSEN.2019.2960323.

    \bibitem{Doppler} C. Noviello, G. Fornaro, and M. Martorella, ``Focused SAR image formation of moving targets based on Doppler parameter estimation,'' \emph{IEEE Transactions on Geoscience and Remote Sensing}, vol. 53, no. 6, pp. 3460--3470, Jun. 2015, DOI: 10.1109/TGRS.2014.2377293.

    \bibitem{Tracking} H. Du, Y. Song, N. Jiang, D. An, W. Wang, C. Fan, and X. Huang, ``A novel SAR ground maneuvering target imaging method based on adaptive phase tracking,'' \emph{IEEE Transactions on Geoscience and Remote Sensing}, vol. 61, art. no. 5211916, Jul. 2023, DOI: 10.1109/TGRS.2023.3294252.

    \bibitem{bessel 1} S. Shi, C. Li, J. Hu, X. Zhang, and G. Fang, ``A high frequency vibration compensation approach for terahertz SAR based on sinusoidal frequency modulation Fourier transform,'' \emph{IEEE Sensors Journal}, vol. 21, no. 9, pp. 10796--10803, May 2021, DOI: 10.1109/JSEN.2021.3056519.

    \bibitem{bessel 2} S. Chen, Y. Wang, and R. Cao, ``A high frequency vibration compensation approach for ultra-high resolution SAR imaging based on sinusoidal frequency modulation Fourier-Bessel transform,'' \emph{Journal of Systems Engineering and Electronics}, vol. 34, no. 4, pp. 894--905, Aug. 2023, DOI: 10.23919/JSEE.2023.000059.

    \bibitem{FrFT} Q. Wang, M. Pepin, A. Wright, R. Dunkel, T. Atwood, B. Santhanam, W. Gerstle, A. W. Doerry, and M. M. Hayat, ``Reduction of vibration-induced artifacts in synthetic aperture radar imagery,'' \emph{IEEE Transactions on Geoscience and Remote Sensing}, vol. 52, no. 6, pp. 3063--3073, Jun. 2014, DOI: 10.1109/TGRS.2013.2269138.

    \bibitem{PGA} D. E. Wahl, P. H. Eichel, D. C. Ghiglia, and C. V. Jakowatz, ``Phase gradient autofocus-a robust tool for high resolution SAR phase correction,'' \emph{IEEE Transactions on Aerospace and Electronic Systems}, vol. 30, no. 3, pp. 827--835, Jul. 1994, DOI: 10.1109/7.303752.

    \bibitem{Sharpness} R. L. Morrison, M. N. Do, and D. C. Munson, ``SAR image autofocus by sharpness optimization: a theoretical study,'' \emph{IEEE Transactions on Image Processing}, vol. 16, no. 9, pp. 2309--2321, Sep. 2007, DOI: 10.1109/TIP.2007.903252.

    \bibitem{Entropy} T. Zeng, R. Wang, and F. Li, ``SAR image autofocus utilizing minimum-entropy criterion,'' \emph{IEEE Geoscience and Remote Sensing Letters}, vol. 10, no. 6, pp. 1552--1556, Nov. 2013, DOI: 10.1109/LGRS.2013.2261975.

    \bibitem{MBSharpness and Sharpness}R. A. Muller and A. Buffington, ''Real-time correction of atmospherically degraded telescope images through image sharpening,'' \emph{J. Opt. Soc. Am.,} vol. 64, no. 9, pp. 1200--1210, Sep. 1974.

    \bibitem{MBSharpness}T. Sakamoto, T. Sato, P. Aubry, and A. Yarovoy, ''Fast imaging method for security systems using ultrawideband radar,'' \emph{IEEE Transactions on Aerospace and Electronic Systems,} vol. 52, no. 2, pp. 658--670, Apr. 2016.

    \bibitem{Sumi} K. Sumi and T. Sakamoto, ``Simulation for noncontact radar-based physiological sensing using depth-camera-derived human 3D model with electromagnetic scattering analysis,''
    arXiv:2507.13826, July 2025. arxiv.org/abs/2507.13826.

    \bibitem{Visualization} T. Shijo, T. Itoh, and M. Ando, ``Visualization of high frequency diffraction based on physical optics,'' \emph{IEICE Trans. Electron.,} vol. E87-C, no. 9, pp. 1607--1614, Sep. 2004.


\end{thebibliography}
\end{document}